\documentclass[11pt,a4paper]{article}
\pdfoutput=1
\usepackage{jheppub}
\usepackage{appendix}
\usepackage[...]{youngtab}

\usepackage{amsfonts}

\catcode`@=11 \@addtoreset{equation}{section} \catcode`@=12

\newcommand{\sid}{\begin{eqnarray*}}
\newcommand{\sidd}{\end{eqnarray*}}
\title{Backreaction effects due to matter coupled higher derivative gravity}
\author{Lata Kh Joshi}
\author{and P. Ramadevi}
\affiliation{Department of Physics,Indian Institute of Technology Bombay,\\
Mumbai 400 076, India}
\emailAdd{latamj@phy.iitb.ac.in}
\emailAdd{ramadevi@phy.iitb.ac.in}
\abstract{AdS-hydrodynamics has proven to be a useful tool  for obtaining transport coefficients observed in
the collective flow of strongly coupled fluids like quark gluon plasma (QGP). 
Particularly, the ratio of shear viscosity to entropy density ${\eta/ s}$  obtained from elliptic flow measurements 
can be matched with the computation done in the dual gravity theory. The experimentally observed temperature dependence of 
${\eta/ s}$ requires the study of scalar matter coupled AdS gravity including higher derivative curvature corrections. 
We obtain the backreaction  to the metric for such a matter coupled AdS gravity in $D$-dimensional spacetime due to the higher 
derivative curvature corrections.  Then, we present the backreaction corrections to shear viscosity $\eta$ and entropy density $s$.}
\begin{document}
\maketitle
 \section{Introduction}
There is a large class of physics problems where the conventional perturbation theory cannot be applied to understand the systems completely. One of the important problem is to understand the dynamics of strongly coupled quantum field theories. This problem has become more interesting after the observation of collective flow exhibited by quark gluon plasma (QGP) in the heavy ion collisions in 
relativistic heavy ion collider (RHIC) and large hadron collider (LHC). 

The introduction of AdS-CFT correspondence \cite{Maldacena:1997re, Witten:1998qj}, has made holographic methods extremely useful to extract possible understanding of strongly coupled systems.
The correspondence allows  supergravity calculations in a $D$-dimensional AdS space with a black brane metric to give the thermal correlators of the strongly coupled
field theories on the $(D-1)$-dimensional boundary.  Various hydrodynamic transport coefficients for the strongly coupled boundary theory can be calculated using real time finite temperature AdS-CFT correspondence \cite{Son:2002sd}. Authors in \cite{Policastro:2001yc,Kovtun:2004de} show that for hydrodynamical boundary systems, the ratio $\eta/s$ of shear viscosity ($\eta$) and entropy density ($s$) has a lower bound of $1/4\pi$. However, for any gauge theory with Einstein gravity dual this bound is saturated giving $\eta/s=1/4\pi$ \cite{Buchel:2003tz}. The violation of the above  bound has also been studied in detail and explained in the pure gravity setting when we add quadratic and higher curvature terms in the gravitational action \cite{Brigante:2007nu, Brigante:2008gz,  Buchel:2008vz, Sinha:2009ev,Kats:2007mq,Camanho:2009vw,Camanho:2010ru,Pal:2009qg,Cremonini:2011iq}. From the string theory point of view, these higher derivative curvature corrections  can be justified as  $\alpha^\prime$  corrections (also known as stringy corrections).  

We know that the finite temperature quantum chromodynamics (QCD) shows conformal invariance only at high temperatures.
In order to study thermal QCD at all temperatures, we require gravity duals describing non-conformal thermal field theories \cite{Gubser:2008yx, Mia:2009wj}. 
One such gravity dual applicable to study thermal QCD involves scalar matter $\Phi$ coupled  to Einstein gravity. In the field theory,  Kubo formula determines  response functions like shear viscosity $(\eta)$ and bulk viscosity $(\xi)$ in terms of the  retarded propagators of stress energy tensors. In the dual gravity side,  adding metric perturbations $\delta g_{i j}$ to the
supergravity solutions leads to the study of these response functions for the boundary quantum field theory. Basically the boundary value of $\delta g_{ij}\equiv \delta g_{ij}(0)$ acts as the source field coupling to the field theory stress tensor $T_{ij}$ \cite{Banerjee:2011tg}. Using the explicit form of the metric solution in the corresponding scalar coupled Einstein gravity theory, we can determine $\eta,\xi$. 
Surprisingly, ${\eta/ s}$ still saturates the bound. This bound gets lowered by the addition of the Gauss-Bonnet terms \cite{Cai:2009zv,Ge:2008ni,Ge:2009eh, Ge:2009ac, Banerjee:2009ju} to the matter coupled Einstein gravity.

The experimental data of average  ${\eta/ s}$ at RHIC energies is different from that of LHC energies \cite{Song:2011qa}, suggesting
that the ratio should be temperature dependent. It is  believed that the plots of ${\eta / s} ~vs ~T$ in the hadronic phase ($T< T_c$) and in deconfining phase ($T> T_c$) increases with the decrease and increase in temperature respectively, where
$T_c$ is the confining-deconfining phase transition temperature. The minima of such a plot will be close to the critical temperature $T_c$. See  Refs.\cite{Song:2011qa,Song:2010mg,Song:2012ua}
for the review on the experimental data fitting to understand ${\eta \over s}(T)$, their upper and lower bounds.

Hence to obtain the temperature dependent ${\eta /s}$ from AdS-hydrodynamics,  the dual gravity action must have higher derivative curvature corrections coupled to matter fields in 
Einstein gravity \cite{Cremonini:2011ej, Cremonini:2012ny}. We will study the gravitational action involving scalar matter \cite{Cremonini:2012ny}. Using AdS-CFT dictionary, the first order perturbative corrections to ${\eta/ s}$ 
depends on the solution for the scalar field $\Phi$.
The form of the scalar field potential $V(\Phi)$ and the profile of $\Phi$  accounts for the temperature dependence  in $\eta/s$. 

The first order computations of ${\eta/s}$ in \cite{Cremonini:2012ny} 
assumes that the spacetime geometry is the solution of matter coupled Einstein gravity. Using the phase variable methods \cite{Gursoy:2008za}
the metric components, expanded near the horizon, can be written in terms of $V(\Phi)$ and its  derivatives. This enables
writing $\eta/s$  for any $V(\Phi)$. They state that the metric corrections (backreaction effects) to the
background geometry, due to higher curvature perturbation, will not add further corrections to
the ratio ${\eta/ s}$ at first order. Hence, the effects of higher derivative corrections on the 
geometry are ignored in their computations.

However, we do see $\eta$ and $s$  getting first order corrections due to backreaction effects
in the geometry of  Einstein gravity with higher derivative curvature terms \cite{Kats:2007mq}. See also 
Ref.\cite{Mia:2009wj} where shear viscosity corrections in the presence of
fundamental matter, which are described by probe branes in gravity duals, are discussed.  
Hence, in this paper we obtain the first order correction (backreaction effects) on the geometry for a matter coupled Einstein gravity with higher derivative curvature terms as perturbations. Then, we obtain the backreaction corrections to shear viscosity and entropy density.  

The plan of the paper is as follows:
In section 2, we will first briefly recapitulate the known $D$-dimensional metric solutions 
of $(i)\colon$Einstein gravity with higher derivative curvature corrections and $(ii)\colon$matter coupled Einstein
gravity, suggesting a convenient form for the metric ansatz. We then obtain the backreaction corrections 
due to higher derivative curvature perturbations to the metric solution.
In section 3, we will  review the holographic method and obtain 
shear viscosity $\eta$ for the spacetime geometry including backreaction corrections.
We will also discuss the Wald entropy formula and evaluate the entropy density $s$ in $D$-dimensions.
Finally, we discuss some of the interesting observations and open problems in the concluding section. 
\section{Black brane metric}
We denote  $D$-dimensional spacetime coordinates as $(t,r,\vec x)$  
where, $\vec x=\{x_1, x_2,..., x_{D-2}\}$, denotes $SO(D-2)$ vector. The $(D-1)$-dimensional boundary 
where field theory resides is obtained by taking $r\rightarrow 0$ limit. In this convention of coordinates,
we assume  scalar field $\Phi$ to be a function of $r$ only. 

We study the metric solutions for the following gravitational action $S$ in $D$-dimensions:

\begin{equation}
\label{action}
S = \frac{1}{2\kappa_D^2} \int{d^{D}x \sqrt{-g} \left(R-\frac{4}{D-2}(\partial \Phi)^2-V(\Phi) + L^2 \beta G(\Phi)R_{\mu \nu \sigma \rho}R^{\mu \nu \sigma \rho}\right)}
\end{equation}
where $L$ is the AdS radius,    $V(\Phi(r))= 2 \Lambda e^{\alpha \Phi(r)}$ (known as Chamblin-Reall potential) with $\Lambda < 0$ and $G(\Phi(r))= e^{\gamma \Phi(r)}$.  Incidentally, for a particular choice of ${\alpha/ \gamma}$, the action in $D$-dimensions can be obtained by $U(1)$ reduction of  
pure gravity action in $D+1$-dimensional string theory. From now on we take  $L=1$.
\subsection{Special limits}
Metric solutions are known for the following limits in the above action $S$:
\begin{itemize}
 \item When we set $\Phi=0$,  $\mathcal{O}(\beta)$ corrections to AdS black brane metric is \cite{Kats:2007mq}
\begin{equation}
\label{kpet}
ds^2 = -\frac{1-r^{D-1}+\beta f(r)}{r^2}dt^2+\frac{dr^2}{r^2(1-r^{D-1}+\beta f(r))}+\frac{1}{r^2}d\vec{x}^2
\end{equation}
where $f(r)$ and $\Lambda$ are
\begin{equation}
f(r)= 2\frac{(D-4)}{(D-2)} + (D-3)(D-4) r^{2(D-1)}~;~
\Lambda= -\frac
{(D-1)(D-2)}{2}
\end{equation}
\item For $\beta=\Phi=0$,  the  metric solution (\ref{kpet}) becomes the familiar
AdS black brane metric.
\item When we set $\beta=0$, the $4$-dimensional solutions are given in Ref.\cite{Kulkarni:2012in} which can be generalised
to $D$-dimensional spacetime as:
\begin{equation}
\label{ansatz}
 ds^2 = -r^{-2a}(1-r^{c})dt^2+\frac{dr^2}{r^{-2a}(1-r^{c})r^4}+r^{-2a}d\vec{x}^2~;~
\Phi(r) = m \text{Log}(r),
\end{equation}
where the forms for $\Lambda,a,c,m$ are :
\begin{eqnarray}
\label{acm}
\Lambda &=&\frac{8 (D-2) \left(-16 (D-1)+(D-2)^2 \alpha ^2\right)}{\left(16+(D-2)^2 \alpha ^2\right)^2}\label{Lamb}~;~
a =\frac{16}{(D-2)^2 \alpha ^2+16},\nonumber\\
c &=& \frac{16 (D-1)-(D-2)^2 \alpha ^2}{(D-2)^2 \alpha ^2+16}~;~
m=\frac{2 (D-2)^2 \alpha }{(D-2)^2 \alpha ^2+16}~.
\end{eqnarray}
\end{itemize}
The solution in the above special limits gives us a clue in proposing an ansatz for metric solution for the action (\ref {action}). 

\subsection{Backreaction corrections}
The eqns.(\ref{kpet}, \ref{ansatz}), representing the two special limits, suggests that we can assume the parameter $a$ 
to be independent of $r,\beta$ but $c$ and $\Phi$ to be dependent on $r,\beta$.  
Hence the ansatz for the complete solution for action (\ref{action}) is,
\begin{equation}
\label{ansatzz}
 ds^2 = -r^{-2a}(1-r^{c(r,\beta)})dt^2+\frac{dr^2}{r^{-2a}(1-r^{c(r,\beta)})r^4}+r^{-2a}d\vec{x}^2~;~
\Phi(r,\beta)~,
\end{equation}
with $a$ given by eqn.(\ref{acm}). Comparing $g_{tt},~g_{rr}$ metric components in the special limits (\ref{kpet}, \ref{ansatz}), we propose
the following form for $c(r, \beta)$: 
\begin{equation}
c(r,\beta) = c+\frac{\text{Log}(1-\beta \kappa(r))}{\text{Log}(r)}~,\label{back}
\end{equation}
where $\kappa(r)$ gives the $\mathcal{O}(\beta)$ back-reaction correction due to the higher derivative curvature term 
in the action (\ref{action}) and $c$ is given by (\ref{acm}). Similarly, we believe the scalar field can also be expanded upto $\mathcal{O}(\beta)$ as
\begin{equation}
\Phi(r,\beta) = m \text{Log}(r)+\beta \Phi_1(r)~,\label{phi}
\end{equation}
where $\Phi_1(r)$ is the $\mathcal{O}(\beta)$ correction to the scalar field and $m$ is as in eqn.(\ref{acm}).  Substituting the ansatz (\ref{ansatzz}) and the proposed forms, (\ref{back},
\ref{phi}) in the Euler-Lagrange equations for the action(\ref{action}) and keeping terms upto $\mathcal{O}(\beta)$ only, gives coupled differential equations. We find exact solutions for $\kappa(r)$ but have not succeeded in finding
$\Phi_1(r)$ from the coupled differential equations. While solving the equations, we had also assumed $\Lambda$ to be $\beta$ dependent. The solutions for $\kappa(r)$ is unchanged even if $\Lambda$ is independent
of $\beta$. Hence, we have taken the same $\Lambda$ as in eqn.(\ref{acm}).

Our main focus in this paper is to find $\mathcal{O}(\beta)$ corrections, $\kappa(r)$, in the metric. This will determine the $\mathcal{O}(\beta)$  shift in the position of  the horizon of the black brane metric which  modifies the first order results of shear viscosity and entropy density. The form of $\Phi_1(r)$ will be needed in the corrections at  second and higher orders in $\beta$. 

We indicate the details of solving the coupled differential equations in the appendix \ref{app-one}. The final result for $\kappa(r)$ 
in $D$-dimensional spacetime turns out to be:
\begin{equation*}
\hspace{0.0in}\kappa(r) =-(D-4)\left(4 +\frac{(D-4) (D-1)  16^{D-\frac{6 \left((-1)^D+1\right) m}{D-2}}(g \times r)^{-a D + 1}}{(D-2) \left((D-2)^2 \alpha ^2-16 (D-1)\right)}\right) +\frac{2 \Lambda  r^{\frac{m}{2} (\alpha +\gamma )}}{(D-2)}\times
\end{equation*}
\begin{equation*}
\hspace{0.1in}\left(r^{m \gamma+a (D-1)}\frac{ \left((D-2)^2\alpha\left(3 (D-2) \alpha+4 (D-3)\gamma \right) -16 (D-4) (D-3)\right)}{ \left((D-2)^2 \alpha  \gamma +16 (D-1)\right)}\right.
\end{equation*}
\begin{equation}
\label{alpha}
\hspace{0.1in}\left.-r^{m \alpha}\frac{\left(4 (D-1) (D-2)^2 \alpha ^2+8 (D-3) (D-2)^2 \alpha  \gamma +64 (D-4)\right)}{ \left((D-2)^2 \alpha  (\alpha +2 \gamma )+16 (D-1)\right)}\right),
\end{equation}
where $g= 16+(D-2)^2\alpha^2$ and $a$, $m$ and $\Lambda$ are given as in eqns.(\ref{acm}).

In the limit when $\alpha$ is zero, the background metric reduces to eqn.(\ref{kpet}). This reinforces that 
the above form of $\kappa(r)$ is consistent and represents $\mathcal{O}(\beta)$ corrections to the spacetime geometry (\ref{ansatz}). 

The  horizon at $r=r_h=1$ (\ref{ansatz}) will  shift  due to  $\mathcal{O}(\beta)$ 
correction $\kappa(r)$ as 
\begin{equation}
\label{horizon}
r_h= 1 + \beta r_1~;~ r_1= \frac{\kappa (r)\vert _{r\rightarrow 1}}{c}.
\end{equation}

The temperature $T$ is given by inverse of  the periodicity in the Euclidean time coordinate $t$. For the above form of the metric, 
$T$ is 
\begin{equation}
\label{temp}
T= \frac{1}{4 \pi} \left.\sqrt{\frac{-g_{tt}^\prime(r)}{g_{rr}^\prime(r)}}\right|_{r \rightarrow r_h}
\end{equation}
where the superscript primes  denote derivatives with respect to $r$.

Now we will perturb this $\mathcal{O}(\beta)$ corrected metric solution by small fluctuations $\delta g_{12}$ where the subscript
represents $x_1,x_2$ directions. The response to this fluctuations in the boundary theory will determine the shear viscosity
$\eta$. We shall present the details of the shear viscosity computations in the following section.

\section{Backreaction effects on shear viscosity and entropy density}
We start this section with a review of shear viscosity in the holographic setting for a strongly coupled boundary theory. We find entropy density using Wald formula for general theories of gravity. We then find  shear viscosity and entropy density for the action (\ref{action}) where we use the full background solution (\ref{ansatzz}) with the known form of $\kappa(r)$  (\ref{alpha}). 
\subsection{Shear viscosity}
Using linear response theory,  the shear viscosity can be determined with the Kubo relation 
\begin{equation}
\label{kubo}
\eta= -\lim_{\omega \rightarrow 0} \frac{1}{\omega} \text{ Im}G^R_{12,12}(\omega, \vec{k}=0),
\end{equation}
where $\omega$ and $\vec k$ denotes the frequency and momentum respectively. The retarded Green's function $G^R_{12,12}(\omega, \vec{k})$ in terms of field theory stress-energy tensor is given by
\begin{equation}
G^R_{12,12}(\omega, \vec{k}) = -i \int{dt d{\vec{x}}e^{i(\omega t- \vec{q}.\vec{x})}\theta(t)\left\langle{\left[T_{12}(x), T_{12}(0)\right]}\right\rangle}
\end{equation}
Thus Kubo formula gives shear viscosity as the low momentum, low frequency limit of the retarded Green's function of stress energy tensor.
We follow the procedure pointed out in \cite{Myers:2009ij} to find the Green's function. In the gravity dual, we  add  metric fluctuations
$\delta g_{12}$ as follows:
\begin{equation}
 g_{12} = g^{(0)}_{12} +\delta g_{12}= 
 g^{(0)}_{12} + \epsilon g_{11}\phi(r, x),~~ {\rm where}~x \equiv (t,\overrightarrow{x})~.\label{fluc}
\end{equation}
In eqn.(\ref{fluc})  $\epsilon$ is the order counting parameter and  
$g^{(0)}_{12}$ is the $12$ component of the unperturbed metric $g$. With the above fluctuations to our metric solution (\ref{ansatzz}), 
we can write the action (\ref{action}) upto $\mathcal{O}(\epsilon^2)$. Substituting the  Fourier transform of $\phi(r,x)$
\begin{equation*}
\phi(r, x)= \frac{1}{(2\pi)^{D-1}} \int{d^{D-1} k~ e^{-i k.x} \phi(r, k)}, ~~{\rm where}~ k\equiv (\omega, \overrightarrow{k}),~~k.x = k_\mu x^\mu,
\end{equation*}
in the action, we obtain $S=S_0+ \epsilon^2 S_{eff}$. The explicit form for $S_{eff}$, usually known as effective action, is
\begin{equation}
S_{eff} \sim \int{\frac{d\omega d^{D-2} \vec k}{(2\pi)^{D-1}} dr \left [A(r)\phi^{\prime\prime}(r,k)\phi(r,-k)+B(r)\phi^\prime(r,k)\phi^\prime(r,-k)\right.}
\end{equation}
\begin{equation*}
\hspace{0.75in} +C(r)\phi^\prime(r,k)\phi(r,-k)+D(r)\phi(r,k)\phi(r,-k)
\end{equation*}
\begin{equation}
\label{action2}
\hspace{0.75in} \left.+E(r)\phi^{\prime\prime}(r,k)\phi^{\prime\prime}(r)+F(r)\phi^{\prime\prime}(r,k)\phi^{\prime}(r,-k)\right]~,
\end{equation}
where $A(r),B(r),\ldots, F(r)$ are functions of $r$ which can be written using metric components and their derivatives. Some of these functions relevant
for the computation of shear viscosity are presented in appendix \ref{app-two}.
The AdS/CFT correspondence dictionary \cite{Witten:1998qj} reads as,
\begin{equation}
\left\langle{\rm {exp}\left(\int{d^{D-1}x~ \phi_0(x)T_{12}(x)}\right)}\right\rangle_{\rm{Field ~ theory}}= \left.{\rm{exp}} \left(-S_{SG}[\phi]\right)\right|_{\rm{Gravity}}
\end{equation}
where, $\phi$ is the solution to the classical supergravity action $S_{SG}[\phi]$ and $\phi_0$, coupling to field theory stress tensor, is the boundary ($r=0$) value of $\phi$. With this relation we can find the two-point correlator of the field theory stress-tensors by using the classical supergravity action. Comparing with Kubo relation, we can read off shear viscosity as \cite{Myers:2009ij}
\begin{equation}
\label{etta}
\eta =\frac{1}{2 \kappa_D^2} \left.\left[ \sqrt{\frac{-g_{rr}}{g_{tt}}} \left( A-B+\frac{F^\prime}{2} \right) +  \left(E \left( \sqrt{\frac{-g_{rr}}{g_{tt}}}\right)^\prime \right)^\prime \right] \right| _{r=r_h}~.
\end{equation}
The explicit form for shear viscosity upto $\mathcal{O}(\beta)$, using $A(r)$, $B(r)$, $E(r)$ and $F(r)$ given in appendix \ref{app-two}, turns out to be
 \begin{equation}
 \label{etaa}
\eta=\frac{1}{2 \kappa^2_D}\left[1-\frac{4\beta (D-2)}{16+(D-2)^2 \alpha ^2}\left(4  r_1-\frac{ \alpha  \gamma (D-2)  \left(16 (D-1)-(D-2)^2 \alpha ^2\right)}{16+(D-2)^2 \alpha ^2}\right)\right]
\end{equation}
where we have used the horizon from eqn.(\ref{horizon}). 
\subsection{Entropy density}
For general theories of gravity with Lagrangian density $\mathcal {L}=\sqrt{-g} L$, Wald formula \cite{Wald:1993nt,Iyer:1994ys} for entropy is: 
\begin{equation}
\label{wald}
\mathcal{S}= -2 \pi \int_{\Sigma}d^{D-2} x \sqrt{-h} \frac{\delta L}{\delta R_{\mu \nu \rho \sigma}} \epsilon_{\mu \nu} \epsilon_{\rho\sigma}~,
\end{equation}
where $h$ is the induced metric at the horizon, $\epsilon_{\mu \nu}$ is binormal to the bifurcation surface, normalized such that $\epsilon_{\mu \nu} \epsilon^{\mu \nu} = - 2$.  
The details of the calculations involving induced metric and binormals are presented in appendix \ref{app-three}. 
The entropy density for action (\ref{action}) upto $\mathcal{O}(\beta)$ is
\begin{equation}
\label{entropy}
s = \frac{1}{2\kappa_D^2}\left[4 \pi -\frac{64 \pi  \beta}{16+(D-2)^2 \alpha ^2}  \left((D-2) r_1-\frac{2 (D-4) \left(16 (D-1)-(D-2)^2 \alpha ^2\right)}{16+(D-2)^2 \alpha ^2}\right)\right]
\end{equation}
These results (\ref{etaa}, \ref{entropy}), in the special limit $\Phi=0$,  agrees with the computations in Ref.\cite{Kats:2007mq}.
\subsection{The ratio $\eta \over s$}
Substituting the backreaction included shear-viscosity and entropy density, we obtain 
the ratio of shear viscosity to entropy density as
\begin{equation}
\label{ratio}
\frac{\eta}{s}=\frac{1}{4 \pi }\left[1-\frac{4\beta  \left(-16 (D-1)+(D-2)^2 \alpha ^2\right) \left(-8 (D-4)+(D-2)^2 \alpha  \gamma \right)}{ \left(16+(D-2)^2 \alpha ^2\right)^2}\right]~.
\end{equation}
Eventhough,  the shift in the horizon $\beta r_1= r_h-1$ (\ref{horizon}) appears in shear viscosity (\ref{etaa}) and entropy density (\ref{entropy}), we 
see the ratio to be $r_1$ independent. 
 
In Ref.\cite{Cremonini:2012ny}, $\eta/s$ and its temperature dependence has been discussed for the action (\ref{action}) in $D=5$ dimensions. They have considered background geometry with no backreaction of higher curvatures. Also, they  apply phase variable method \cite{Gursoy:2008za} to write $\eta/s$ in terms of the scalar potential, $V(\Phi(r))$ and coupling $G(\Phi(r))$ giving
the following form:
\begin{equation}
\frac{\eta}{s}= \frac{1}{4\pi}\left[1+\frac{2}{3}\beta\left(G(\Phi(r_h))V(\Phi(r_h))-\frac{9}{8}G^\prime(\Phi(r_h))V^\prime(\Phi(r_h))\right)\right]~. \label{ratio1}
\end{equation}
Note that, the action in Ref.\cite{Cremonini:2012ny} can be brought to the form (\ref{action}) with a suitable scaling of scalar field $\Phi(r)$
leading to the above result.
Substituting $V(\Phi), G(\Phi)$ in the above equation gives eqn.(\ref{ratio}) when $D=5$. 
From the equality of ratios (\ref{ratio}, \ref{ratio1}) at $\mathcal{O}(\beta)$, we believe the linear backreaction effects  in the $\eta/s$ will appear at higher orders of $\beta$. 

The reason for the $r_1$ independence in $\eta/s$ is not obvious from our calculations at horizon $r_h=1+\beta r_1$. In order to understand the
$r_1$ independence we have presented the forms of $\eta$ and $s$ at $r_h=r_0+ \beta \tilde r_1$ in appendix \ref{app-four}. The presence of $r_0$ is crucial to mathematically see the statement given in footnote $8$ of Ref.\cite{Cremonini:2012ny}.

\section{Conclusion}
In this work, we considered higher derivative curvature terms added to  matter coupled Einstein gravity action. Our main aim was to find the complete solution for such an action.
Treating higher derivative curvature terms as perturbations with coupling $\beta$, we succeeded in finding $\mathcal{O}(\beta)$ corrections (backreaction effects)
to the $D$-dimensional spacetime metric. The horizon shifts from $r=1$ to $r=1+\beta r_1$(\ref{horizon}) due to backreaction effects.

With the complete spacetime geometry including backreaction effects available, we  revisited the first order corrections to shear viscosity $\eta$  and entropy density $s$.
Both of them have the horizon shift  dependence. For any $D$-dimensions, the $r_1$ dependence cancels in the ratio $\eta/s$ . Though this 
result is expected from the statement in Ref.\cite{Cremonini:2012ny}, we provide a detailed mathematical proof in appendix \ref{app-four}.

Going beyond first order is  a  challenging mathematical problem which needs to be studied.
One of the barriers we faced in going to second order in $\beta$  is the inablility to find solution for $\Phi_1(r)$. Probably,
some other form of metric ansatz or numerical methods have to be explored to find $\Phi_1(r)$. We would also like to determine backreaction corrections for other
gravity theories like Einstein-Maxwell-dilaton gravity with higher derivative curvature corrections \cite{Cremonini:2011ej}.

We have done our analysis for a fixed potential $V(\Phi(r))= e^{\alpha \Phi(r)}$ upto  $\mathcal{O}(\beta)$. 
Phenomenologically this potential is important only at large $\Phi$ \cite{Gubser:2008yx}. 
Ideally, we should solve background geometry for a phenomenologically relevant potentials and reproduce
the experimental data on $\eta/s$.  Our methodology still faces  mathematical difficulty in solving Euler-Lagrange
equations. We need radical ideas to tackle such difficulties and find complete solutions for these potentials. We hope to pursue these 
problems in future.

\section*{Acknowledgement}
We would like to thank A. K. Mohanty and S.V. Suryanarayana for discussions on experimental data of QGP. We thank Keshav Dasgupta, Mohammed Mia,
Krishna Rajagopal and Umut Gursoy for the various helpful discussions on the topic. We also thank Utkarsh Sharma for reading the manuscript and 
giving valuable comments.

\begin{appendix}
\section{Back-reaction on the background}
\label{app-one}
Let us rewrite the action (\ref{action}) as,
\begin{equation}
S=\frac{1}{2\kappa_D^2} \int{d^{D}x \sqrt{-g} (R+L^{matter}+L^{hd})}
\end{equation}
with
\begin{eqnarray}
L^{matter} =  -\frac{4}{D-2}(\partial \Phi)^2-V(\Phi)~;~
L^{hd} =  \beta G(\Phi)R_{\mu \nu \sigma \rho}R^{\mu \nu \sigma \rho}.
\end{eqnarray}
Varying the action with respect to  scalar field $\Phi(r)$  gives the following  Euler-Lagrange equation:
\begin{eqnarray}
\frac{8}{D-2} \Box \Phi +\frac{\partial}{\partial \Phi } \left[\beta  R_{\mu \nu \sigma \rho}R^{\mu \nu \sigma \rho} G(\Phi )-V(\Phi )\right]=0.\label{eom2}~.
\end{eqnarray}
Similarly, varying with respect to metric field $g^{\mu\nu}$ gives  Einstein's equations: 
\begin{equation}
R_{\mu \nu}-\frac{1}{2} g_{\mu \nu}R = T_{\mu \nu}^{matter}+T_{\mu \nu}^{hd} \label{eom1}~,
\end{equation}
where
\begin{align}
T_{\mu \nu}^{matter}=&\frac{1}{2}g_{\mu \nu} L^{matter}~,\\
T_{\mu \nu}^{hd}=&2 \beta G(\Phi) \left(g_{\mu \nu}R_{\rho \sigma \kappa \delta}R^{\rho \sigma \kappa \delta}-4 R_{\mu \sigma \kappa \delta}R_\nu^{\sigma \kappa \delta}-8 \Box R_{\mu \nu}\right.\nonumber\\
&\hspace{1in}\left.+4 \nabla_\nu \nabla_\mu R+8 R^\rho_\mu R_{\rho\nu}-8 R^{\rho \sigma}R_{\mu \rho \nu \sigma }\right)~.
\end{align}
For the metric ansatz with diagonal metric components and possessing $SO(D-2)$ symmetry in $\vec x$ coordinates, 
the number of independent equations to solve in eqn.(\ref{eom1}) are only three.

Substituting the ansatz (\ref{ansatzz}) in the eqns. (\ref{eom2}, \ref{eom1}), we have  to solve four coupled differential equations
to determine $\kappa(r), \Phi_1$ which are the $\mathcal{O}(\beta)$  corrections. Taking a suitable linear combination of Einstein equations (\ref{eom1}) 
 for  metric components $g^{tt}$ and $g^{rr}$, we find a 
differential equation for $\kappa(r)$ which is a solvable equation. The solution for $\kappa(r)$  can be written
in $D$-dimensions as given in eqn.(\ref{alpha}). Once $\kappa(r)$ is known, we obtain a differential equation for $\Phi_1$ and we have not
managed to find its explicit form.
\section{Coefficients used in $\eta$}
\label{app-two}
The coefficients in eqn.(\ref{action2}) used in the calculation of shear viscosity for action (\ref{action}) are,
\begin{align}
A(r)=&\frac{2 \sqrt{g_{tt}(r)} g_{11}(r){}^{\frac{D}{2}-3}}{g_{rr}(r){}^{5/2}} \left[\beta  e^{\gamma  \phi (r)} \left\{g_{11}(r) \left(g_{rr}'(r) g_{11}'(r)\right.\right.\right.\nonumber\\
&\left.\left.\left.-2 g_{rr}(r) g_{11}''(r)\right)+g_{rr}(r) g_{11}'(r){}^2\right\}+g_{rr}(r){}^2 g_{11}(r){}^2\right]\\
B(r)=&\frac{g_{11}(r){}^{\frac{D}{2}-3}}{2 g_{tt}(r){}^{3/2} g_{rr}(r){}^{7/2}} \left[g_{tt}(r){}^2 \left\{\beta  e^{\gamma  \phi (r)} \left((D+1) g_{rr}(r){}^2 g_{11}'(r){}^2\right.\right.\right.\nonumber\\
&\left.\left.\left.-2 g_{rr}(r) g_{11}(r) \left(g_{rr}'(r) g_{11}'(r)+2 g_{rr}(r) g_{11}''(r)\right)+g_{11}(r){}^2 g_{rr}'(r){}^2\right)\right.\right.\nonumber\\
&\left.\left.+3 g_{11}(r){}^2 g_{rr}(r){}^3\right\}+\beta  g_{rr}(r){}^2 g_{11}(r){}^2 g_{tt}'(r){}^2 e^{\gamma  \phi (r)}\right]\\
E(r)= &\frac{2 \beta  \sqrt{g_{tt}(r)} g_{11}(r){}^{\frac{D}{2}-1} e^{\gamma  \phi (r)}}{g_{rr}(r){}^{3/2}}\\
F(r)=&\frac{2 \beta  \sqrt{g_{tt}(r)} g_{11}(r){}^{\frac{D}{2}-2}}{g_{rr}(r){}^{5/2}} \left[2 g_{rr}(r) g_{11}'(r)-g_{11}(r) g_{rr}'(r)\right] e^{\gamma  \phi (r)}
\end{align}
The $g_{\mu \nu}$ in these functions are the background metric components obtained from $ds^2=g_{tt}(r)dt^2+g_{rr}(r)dr^2+ g_{11}(r)\sum_{i,j=1}^{D-2} \delta_{ij} dx^i dx^j$.
\section{Entropy density (s)}
\label{app-three}
For general $D$-dimensional gravity theories described by a Lagrangian density $\mathcal{L}=\sqrt{-g}L$, Wald has proposed a formula for entropy $\mathcal S$ (\ref{wald}). The
entropy density $s$, which gives the degrees of freedom per unit volume residing on the horizon $r=r_h$ ($D-2$-dimensional hypersurface), is 
\begin{equation}
\label{ed}
s= -2 \pi \sqrt{-h} \frac{\delta L}{\delta R_{\mu \nu \rho \sigma}} \epsilon_{\mu \nu} \epsilon_{\rho\sigma},
\end{equation}
where $h$ is the determinant of the induced metric on the $(D-2)$-dimensional hypersurface. We can choose the binormals $\epsilon_{\mu \nu}$ as \cite{Astefanesei:2008wz}
$$
\epsilon_{\mu \nu} = n_\mu k_\nu - n_\nu k_\mu
$$
where $n$ and $k$ are the null vectors normal to the bifurcate Killing horizon, with $n . k = 1$.
One choice for these null vectors, $n$ and $k$ can be:
\begin{eqnarray}
n= \frac{\partial}{\partial t}~~;~~ k=-\frac{1}{g_{tt}(r)}\frac{\partial}{\partial t}-\frac{\partial}{\partial r}. 
\end{eqnarray}
For the action (\ref{action}), using the form of $L$  
$$L={1 \over 2\kappa_D^2}\left[R-\frac{4}{D-2}(\partial \Phi)^2-V(\Phi) + \beta G(\Phi)R_{\mu \nu \sigma \rho}R^{\mu \nu \sigma \rho}\right] ~,$$ 
gives
\begin{equation}
\frac{\partial L}{\partial R_{abcd}}  \epsilon_{ab} \epsilon_{cd} ={1 \over 2\kappa_D^2}\left[ -2-4\beta G(\Phi)(R-2h^{ab}R_{ab}+h^{ac}h^{bd}R_{abcd})\right].
\end{equation}
The induced metric $h_{ab}$ can be obtained using the following relation between the original and induced metric,
\begin{equation}
h_{ab}= g_{ab}-(n_a k_b-n_b k_a).
\end{equation}
Using these equations, we get the entropy density upto $\mathcal{O}(\beta)$ as given in eqn.(\ref{entropy}).
\section{$\eta$ and $s$ for any horizon}
\label{app-four}
It is intriguing to see that the $\mathcal O(\beta)$ corrections to $\eta$ and $s$, due to horizon shift, does not alter $\eta/s$ result. This result appears to be a puzzle at least from the formal calculations done with the AdS horizon taken as $r_h=1$. To explicitly see what causes $\eta/s$ to be unaffected due to backreaction, we will take the AdS horizon as $r_h=r_0$. The complete metric solution with $\mathcal O(\beta)$ backreaction included will be
\begin{equation}
ds^2 = -r^{-2a}\left(1-\left(\frac{r}{r_0}\right)^{\tilde{c}(r,\beta)}\right)dt^2+\frac{dr^2}{r^{-2a}\left(1-\left(\frac{r}{r_0}\right)^{\tilde{c}(r,\beta)}\right)r^4}+r^{-2a}d\vec{x}^2,
\end{equation}
where $\tilde{c}(r, \beta)|_{r_0=1}= c(r,\beta)$ (\ref{back}) and $a$ is given as in eq.(\ref{acm}). Formally, we can write the new horizon as $r_h= r_0+ \beta \tilde{r}_1$. Using the above data for complete metric solution, we obtained the following results for $\eta$ and $s$:
\begin{eqnarray}
\eta&=&\frac{1}{2 \kappa_D^2}\left[r_h^P-\frac{4 \beta  (D-2)^2 \alpha  \gamma  \left(\alpha^2(D-2)^2 -16 (D-1)\right) r_0^M}{\left(16+\alpha^2(D-2)^2 \right)^2}\right]\label{appeta}\\
s&=&\frac{1}{2 \kappa_D^2}\left[4 \pi  r_h^P-\frac{128 \pi  \beta  (D-4) \left(\alpha^2 (D-2)^2 -16 (D-1)\right) r_0^N}{\left(16+\alpha^2(D-2)^2\right)^2}\right]\label{apps}
\end{eqnarray}
with
\begin{eqnarray}
P&=&-\frac{16 (D-2)}{16+\alpha^2(D-2)^2 },\nonumber\\
M&=&\frac{2 \left(-8 (D-2)+(D-2)^2 \alpha  \gamma+(D-2)^2 \alpha^2 \right)}{16+\alpha^2(D-2)^2},\nonumber\\
N&=&\frac{2 \left(8 (D+2)+(D-2)^2 \alpha  \gamma+(D-2)^2 \alpha ^2 \right)}{16+\alpha^2(D-2)^2 }.
\end{eqnarray}
The above expressions (\ref{appeta}, \ref{apps}) agree with the eqns. (\ref{etaa}, \ref{entropy}) at $r_0=1$ where $\tilde{r}_1=r_1$. We observe both $\eta$ and $s$ have leading term proportional to $r_h^P$. Incorporating the backreaction effects, the leading term will give the corrections to $\eta$ and $s$ due to horizon shift. That is, $r_h= r_0+\beta \tilde{r}_1$ will give $\tilde r_1$ corrections. Taking the ratio $\eta/s$, we see that the $\tilde{r}_1$ dependence cancels.

In the absence of higher derivative terms ($\beta=0$), the gravity theory we considered must obey $\eta/s=1/4\pi$ universality bound. For such theories, the leading term in $\eta$ and $s$ must have the same powers of $r_h$. The horizon shift will not affect the ratio $\eta/s$ for these theories which is consistent with the footnote $8$ in Ref.\cite{Cremonini:2012ny}.

We have also looked at charged matter gravity theories \cite{Cremonini:2011ej} where the $\eta/s$ bound is violated when we switch off the higher derivative terms. Even though first order backreaction to metric is not known, we can use the above argument to justify that the ratio $\eta/s$ will be independent of the horizon shift.
\end{appendix}
\bibliographystyle{JHEP}
\bibliography{biblio}
\end{document}